\documentclass{article}

\PassOptionsToPackage{numbers, compress}{natbib}
\usepackage[final]{neurips_2023_ml4ps}

\usepackage[utf8]{inputenc} 
\usepackage[T1]{fontenc}    
\usepackage{url}            
\usepackage{booktabs}       
\usepackage{amsfonts}       
\usepackage{nicefrac}       
\usepackage{microtype}      
\usepackage{xcolor}         
\usepackage{graphicx}
\usepackage{amsmath}
\usepackage{amssymb}
\usepackage{mathtools}
\usepackage{amsthm}
\usepackage[capitalize,noabbrev]{cleveref}

\title{Discovering Black Hole Mass Scaling Relations with Symbolic Regression}

\author{%
  Zehao Jin \\
  Center for Astrophysics and Space Science (CASS)\\
  New York University\\
  Abu Dhabi, PO Box 129188, Abu Dhabi, UAE \\
  \texttt{zj448@nyu.edu} \\
  \And
  Benjamin L. Davis \\
  Center for Astrophysics and Space Science (CASS)\\
  New York University\\
  Abu Dhabi, PO Box 129188, Abu Dhabi, UAE \\
  \texttt{bld5865@nyu.edu} \\
}

\begin{document}

\maketitle

\begin{abstract}
Our knowledge of supermassive black holes (SMBHs) and their relation to their host galaxies is still limited, and there are only around 150 SMBHs that have their masses directly measured and confirmed.
Better black hole mass scaling relations will help us reveal the physics of black holes, as well as predict black hole masses that are not yet measured.
Here, we apply symbolic regression, combined with random forest to those directly-measured black hole masses and host galaxy properties, and find a collection of higher-dimensional (N-D) black hole mass scaling relations.
These N-D black hole mass scaling relations have scatter smaller than any of the existing black hole mass scaling relations.
One of the best among them involves the parameters of central stellar velocity dispersion, bulge-to-total ratio, and density at the black hole's sphere-of-influence with an intrinsic scatter of $\epsilon=0.083\,\ \textup{dex}$, significantly lower than $\epsilon \sim 0.3\,\textup{dex}$ for the M-$\sigma$ relation.
These relations will inspire black hole physics, test black hole models implemented in simulations, and estimate unknown black hole masses on an unprecedented precision.
\end{abstract}

\section{Introduction}
\label{sec:intro}

Supermassive black holes (SMBHs) are some of the most extreme objects in the Universe, and are incredibly intriguing and captivating.
Yet, we know little about black holes, both in terms of theory and observation.
Theoretically, there is no agreement on how an SMBH itself evolves and how it affects the evolution of its host galaxy (i.e., "so-called" SMBH feedback).
Observationally, we only had our first “photo” of an SMBH from the EHT \citep[Event Horizon Telescope;][]{Akiyama_2019} a few years ago, and until now, only $\sim$150 SMBHs have had their mass directly measured through the dynamics of nearby objects. 

One of the best ways to understand black holes is through black hole mass scaling relations – an empirical correlation between a central black hole mass and a property of its host galaxy.
The M--$\sigma$ (black hole mass – velocity dispersion) relation is the most prominent and studied black hole mass scaling relation \citep{Kormendy_2013}.
Since a black hole is evolving together with its host galaxy, the black hole mass will not only correlate with $\sigma$, but also correlate with many other galaxy properties, such as mass, luminosity, radius, density, velocities, morphology, galaxy type, etc.
In fact, we do find many black hole mass scaling relations from these galaxy properties \citep[e.g.,][]{Davis_2018,Sahu_2019,Sahu_2020,Graham_2022}.
It is then natural to ask which galaxy properties are intrinsically connected to the central black hole mass, and which galaxy properties are simply degenerate with other ones.
To shed light on this question, one can look for N-D (multi-dimensional) black hole mass scaling relations, which relates black hole mass to multiple galaxy properties in one equation.
Such equations should comes in two tracks, for unveiling physics and for making predictions, respectively:

\textbf{Low-scatter track:} The equation that reveals the nature of physics should by construction give precise predictions (i.e., have low scatter in terms of scaling relations), and at the same time, stay concise such that it can be interpreted by physicists.
Finding as low scatter as possible, but as simple as possible black hole mass scaling relations, will reveal which and how galaxy properties are intrinsically connected to the central black hole.
These relations can also be used to test the AGN (active galactic nucleus) models implemented in modern cosmological simulations such as NIHAO AGN \citep{blank_2019}, IllustrisTNG \citep{Nelson_2017}, and FIRE \citep{Onorbe_2015}.

\textbf{Easy-to-use track:} Only around 150 very nearby black holes have their mass directly measured, while there are billions of galaxies out there in the Universe, and the mass of their central black holes are not directly measurable.
Although a ``low-scatter" relation gives the best prediction, one might sometime lack certain key properties to use such a relation, especially for high-redshift galaxies where not all galaxy properties can be easily measured.
To best predict black hole masses, there is a need for a set of ``easy-to-use'' relations that can make the most precise prediction possible of black hole mass, given a certain limited set of measurable galaxy properties.

Symbolic Regression (SR) is a perfect tool to look for both these two tracks of scaling relations.
SR searches for the best equation that fits a given data set, and has already powered the discovery or re-discovery of many empirical relations, as in \cite{Krone_Martins_2014,cranmer_2019,cranmer_2020,lemos_2022}.
In this extended abstract, we will apply SR to the black hole mass -- galaxy property catalog to discover a set of ``low-scatter" track and ``easy-to-use" track relations that significantly outperform any current 2-D black hole mass scaling relation.
The dimensionality of the catalog allows us to conduct an extensive SR equation search over a large number of features ($\sim$70 different galaxy properties), while on limited entries of data ($\sim$100 galaxies).

\section{Method: Symbolic Regression}
\label{sec:method}

One of the best ways to find both ``low-scatter” relations and ``easy-to-use” relations is to use SR.
SR is a sub-field of machine learning that aims to find mathematical expressions that best fit a given set of data.
Given a target parameter, a set of potentially relevant parameters, and a pool of operators, SR will come up with equations that best recovers the target parameter using a certain selection of relevant parameters, constants, and operators.

SR is a war between precision and simplicity: by Occam’s razor, the desired physics equation or relation should explain the phenomena (have a low scatter), while maintaining the beauty of simplicity (i.e., use a minimal number of parameters, constants, and operators).
SR's precision vs.\ simplicity nature is exactly what the ``low-scatter" relation, or the relation that reveals physics is looking for, as described in \cref{sec:intro}.
On the other hand, SR by construction is the perfect way to find the ``easy-to-use" relations --- simply pass the set of observable galaxy properties to SR, and SR searches for the relation that gives the best black hole mass prediction.

Specifically, we adopt the \textsc{\small{Python}} SR package \textsc{\small{PySR}} \citep{cranmer_2023}, a multi-population evolutionary algorithm to search for equations given a variable pool and an operator pool.
During the search, \textsc{\small{PySR}} judges proposed equations with a score that aims to maximize the accuracy and penalize the complexity, and the balance between accuracy and complexity is regulated by a parsimony constant.
The accuracy is simply the mean squared error loss in this work, and the simplicity is characterized by a complexity score based on the number of variable, operators, and constants involved in the equation.
The operator pool in this work is simply $+$, $-$, $\times$, and $\div$ since most of the variables enter \textsc{\small{PySR}} in their $\log_{10}$ form, but we also test the cases with additional $\log_{10}$, power, and exponentiation included for exploration. 
We will discuss more on the choice of operators and the handling of input variables with inductive bias in a coming full paper. 
Note that \textsc{\small{PySR}} does not deterministically nor uniquely find the best equation, but comes up with a list of equation candidates from the infinite set of possible ones, with a reasonably good combination of precision and simplicity \citep{lemos_2022}.

\begin{figure}
  \centering
  \includegraphics[width=\linewidth]{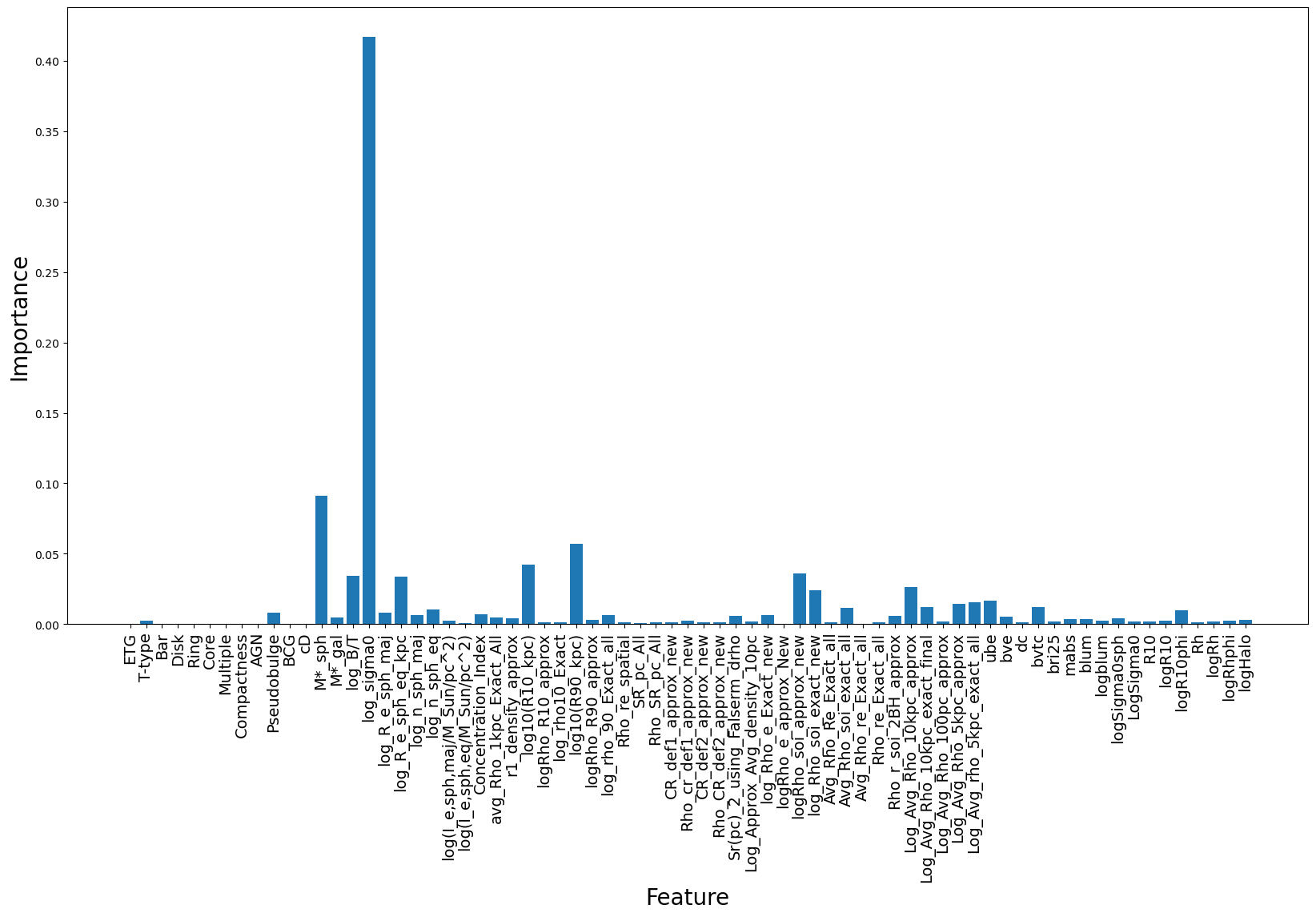}
  \caption{Random forest feature importance of 71 galaxy properties studied for the ``low-scatter track'' relations.}
  \label{fig:rf_low-scatter}
\end{figure}

\begin{figure}
  \centering
  \includegraphics[width=\linewidth]{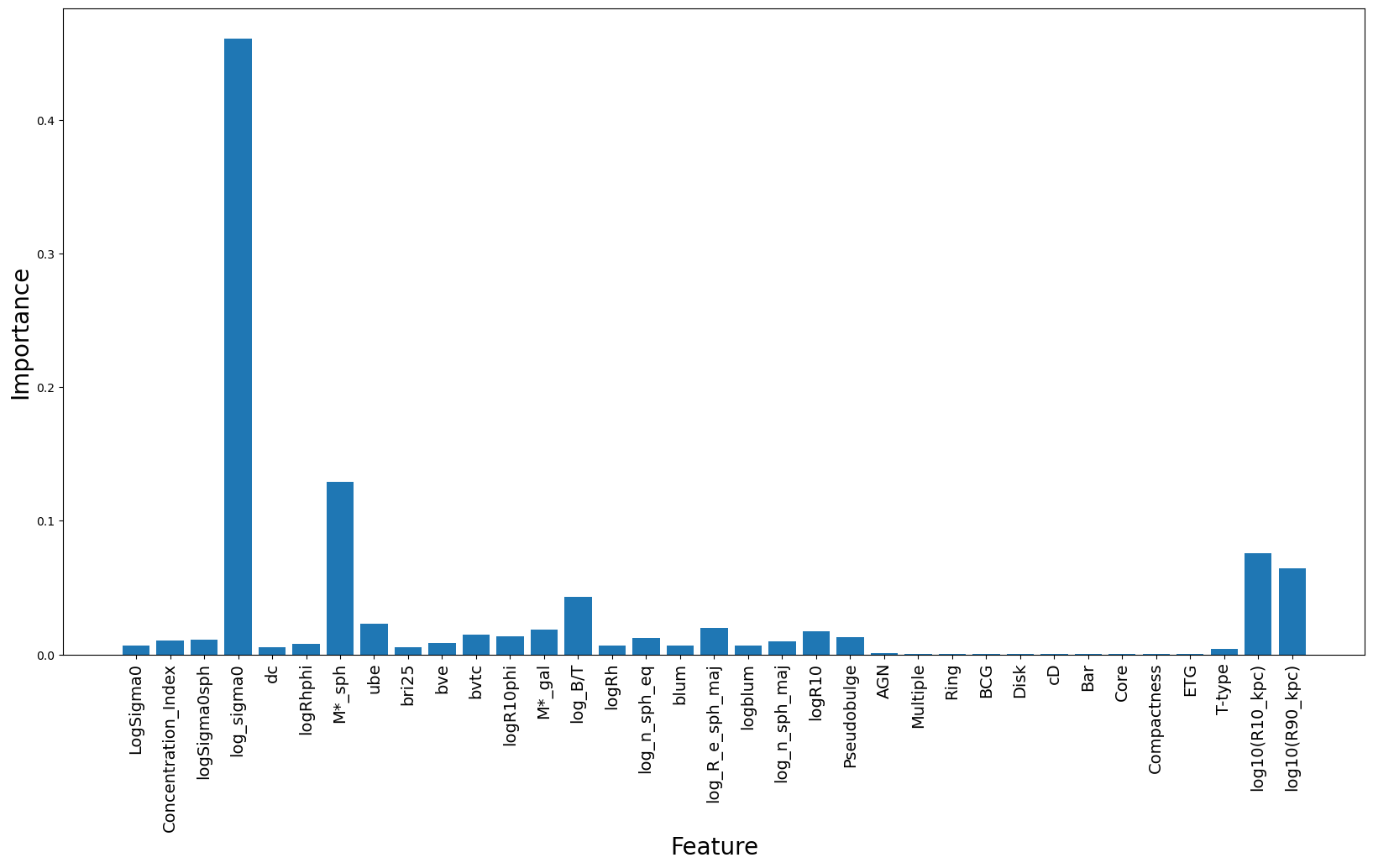}
  \caption{Random forest feature importance of 35 galaxy properties studied for the ``easy-to-use track'' relations.}
  \label{fig:rf_easy-to-use}
\end{figure}

The variable pool is determined with the help of random forest. Random forest can be used for regression tasks, and at the same time, gives the importance for each of the input variables that leads to the regression prediction. 
We first run a random forest regression to predict black hole mass with all variables available, and pick the top 10 important variables as our initial variable pool for SR (see \cref{fig:rf_low-scatter} \& \cref{fig:rf_easy-to-use} for feature importance maps). After the initial search, we mutate the variable pool based on our prior knowledge of physics to test potentially promising variables and run SR repeatedly.


\section{Data: the Black Hole Mass –- Galaxy Property Catalog}
\label{sec:data}

The black hole mass –- galaxy property catalog contains 145 direct measurements of black hole masses, plus more than 100 (still growing) different properties of their host galaxies, as well as the measurement error in each quantity.
Typical galaxy properties in the catalog include velocity ($\sigma$, $v_{\textup{rot}}$, …), mass ($\text{M}^*_{\text{sph}}$, $\text{M}^*_{\text{gal}}$, $\text{M}^*_{\text{DM}}$, $B/T$,  …), radius ($R_{e}$, $R_{10}$, $R_{90}$, $R_\text{soi}$, …), density ($\rho_{e}$, $\rho_{10}$, $\rho_{90}$, $\rho_\text{soi}$, …), luminosity/color ($L_B$, $L_V$, $bvc$, …), morphological classification (ETG/LTG, T-type, Bar, Disk, Core, Pseudobulge, … ), etc\footnote{$B/T$: Bulge to total mass ratio, i.e., $\frac{\text{M}^*_{\text{sph}}}{\text{M}^*_{\text{gal}}}$; soi: sphere-of-influence; $bvc$: B-V color; morphology class are Boolean}.
This project is hosted in a public GitHub repository\footnote{\scriptsize{\url{https://github.com/ZehaoJin/Ultimate_black_hole_mass_scaling_relations_Symbolic_Regression}}}, where you can also find the black hole mass –- galaxy property catalog. 

The aim of this project is to examine scaling relations between a multitude of physical galaxy parameters against a sample of ``directly-measured'' supermassive black hole (SMBH) masses.
As such, we restrict our sample to only host galaxies that have had their central black holes measured via first principles.
We consider direct measurements to be only those that can observe motions within the sphere-of-influence of an SMBH.
Curation of this sample of galaxies began a decade ago and has continued to grow as new direct measurements have been added to the literature \citep{Graham_2013,Savorgnan_2016,Davis_2017,Davis_2019,Davis_2019b,Sahu_2019,Sahu_2019b}.
The sample currently stands at 145 galaxies, with new galaxies added as the inventory of direct measurements grows.

The search space and computational cost of genetic SR algorithms such as \textsc{\small{PySR}} scales with both the size of the variable pool, the size of the operator pool, and the size of data. 
Although in the black hole mass –- galaxy property catalog, the number of variables is very high, the number of data is relatively low, which made SR possible.
Besides, common machine learning attempts such as neural networks, or even decision tree based methods like random forest are very likely to over-fit given such a high-dimensional but low-number dataset, are thus less powerful to extrapolate and predict black hole masses compared to a concise relation. 
While in terms of interpretability, nearly nothing compares to an exact mathematical expression.

This particular work aims to find scaling relations that work for all kinds of morphologies of the host galaxies, whereas a previous work \citep{Davis_Jin_2023} focus on the spiral galaxies within this catalog and presents a planar relation that yields the best-in-class accuracy in prediction of black hole masses in spiral galaxies.

\section{Results}
\label{sec:result}

\begin{table*}[t]
\tiny
\caption{N-D black hole mass scaling relations found by SR, compared to 2-D relations.
All relations give black hole mass in $\log (\frac{\text{M}_\text{BH}}{\text{M}_\odot})$.}
\label{tab:relations}
\vskip 0.15in
\begin{center}
\begin{tabular}{llll}
\toprule
\textsc{N-D Relation} & \textsc{2-D Counterpart} & \textsc{N-D RMSE} & \textsc{2-D RMSE} \\
\toprule
\textbf{Low-scatter relations} \\
\midrule
$2.85 \log \left(\frac{\sigma_0}{189}\right) + 1.16 \log \left(\frac{B/T}{0.437}\right) - 0.33 \log \left(\frac{\rho_\text{soi}}{601}\right) + 8.20 $ 
& $6.10 \log \left(\sigma_0/200\right)+8.27$
&  0.27 & 0.46 \\

& $2.40 \log(B/T) + 9.11$
&  & 0.57 \\

& $-0.93 \log \rho_\textup{soi} + 10.64$
&  & 0.71 \\

\hline

$2.57 \log \sigma_0 + 0.38 (\log R_{e,sph,eq} - \log \rho_\textup{soi}) + bvc + 2.61$
& $6.10 \log (\sigma_0/200)+8.27$
& 0.26 & 0.46 \\

\hline

$2.57 \log \sigma_0 + \log(B/T) - 0.43 \log \rho_\textup{soi} - 0.24 \ \textup{Pseudobulge} + 4.01$
& $6.10 \log (\sigma_0/200)+8.27$
& 0.23 & 0.46 \\

\toprule
\textbf{Easy-to-use relations} \\
\midrule

$\log \sigma_0 + \log \textup{M}^{*}_{\textup{sph}} - 0.56 \ \textup{Pseudobulge} - 4.57$
& $1.31 \log \text{M}^*_\textup{sph} - 5.83$
& 0.33 & 0.43 \\

\hline

$0.93 (\log \sigma_0)^2 + 0.56 \log R_{e,sph,maj} + 3.2)$
& $1.51 \log R_{e,sph,maj} + 7.61$
& 0.34 & 0.59 \\

\hline

$3.59 \log \sigma_0 + 0.50 \log R_{e,sph,maj} - 0.50 \textup{Pseudobulge}$
& $6.10 \log (\sigma_0/200)+8.27$
& 0.30 & 0.46 \\

\bottomrule
\end{tabular}
\end{center}
\vskip -0.1in
\end{table*}

\subsection{Low-Scatter Track relations}
\label{sec:low-scatter}

After repeatedly running SR with the variable pool guided by random forest and prior physics knowledge, we came up with a set of ``low-scatter" relations presented in the first half of Table \ref{tab:relations}.
All of these relations carry a significantly lower scatter than any of the existing relations.
Due to the limited length of this abstract, we are not able to further prove if these ``low-scatter" relations are intrinsically concise, i.e., the introduction of any extra dimension of galaxy property is statistically worth the improvement in scatter that an extra dimension brings in, and any particular galaxy property is interchangeable with another one while keeping similar scatter.
The physics behind any of these relations is also beyond the scope of this abstract, and we will leave it for future work.

When applying \textsc{\small{PySR}}, we assign weight to each entry according to the measurement error in black hole mass. To also consider the measurement error in galaxy properties, get uncertainty on constants, and arrive an intrinsic scatter\footnote{Intrinsic scatter is the RMSE when there is zero measurement error.}, we first normalize each variable with its median, and perform an MCMC to re-fit the formula obtained by \textsc{\small{PySR}} with \textsc{\small HyperFit} \citep{Robotham_2015}. In the current work, this additional step is only conducted on the the linear $\sigma_0$--$(B/T)$--$\rho_\text{soi}$ relation\footnote{In units of $\sigma_0 [\textup{km/s}]$ and $\rho_\text{soi} [\textup{M}_\odot \cdot \textup{pc}^{-3}]$} ($1^{st}$ row of \cref{tab:relations})

\begin{equation}
\label{equ:best}
\scriptsize
\log \left(\frac{\text{M}_\text{BH}}{\text{M}_\odot}\right)=\alpha \log \left(\frac{\sigma_0}{189}\right) + \beta \log \left(\frac{B/T}{0.437}\right) + \gamma \log \left(\frac{\rho_\text{soi}}{601}\right) + \delta
\end{equation}

with $\alpha=2.854\pm0.035$, $\beta=1.162\pm0.046$, $\gamma=-0.334\pm0.026$, $\delta=8.203\pm0.036$, and with an intrinsic scatter $\epsilon=0.087\pm0.057 \ \textup{dex}$ in the $\log (\textup{M}_\textup{BH}/\textup{M}_\odot)$ direction from a sample of 122 galaxies\footnote{122 out of 145 galaxies have all of $\sigma_0$, $B/T$, and $\rho_\text{soi}$ measured. The \textsc{\small HyperFit} re-fit is conducted on these 122 galaxies.}.  
An $\epsilon=0.087\ \textup{dex}$ is far better than an $\epsilon=0.29\ \textup{dex}$ in a sample of 75 galaxies \cite{Kormendy_2013}, or an $\epsilon=0.43\ \textup{dex}$ in a sample of 137 galaxies \citep{Sahu_2019}, obtained by the M--$\sigma$ relation.

We also judge some of our relations with commonly adopted model selection criteria, the Akaike Information Criterion (AIC) and the Bayesian Information Criterion (BIC), where the model accuracy and complexity are balanced in such criteria. The new relations found by \textsc{\small{PySR}} outperforms all existing 2-D relations.
As \cref{tab:relations} and \cref{tab:AICBIC} shows, any of $\sigma_0$, $B/T$, and $\rho_\text{soi}$ alone will not give the low scatter obtained by all three together.
Similarly, although there is no space to show in this abstract, any combination of only two from the three parameters will not result in same level of scatter obtained by all three.
All of the three parameters alone are well motivated physically as described in literature \citep{Kormendy_2013,Davis_2018,Sahu_2022}, but this relation connects the three parameters together, form the very center of a galaxy to larger-scale galaxy morphology, to inspire a more complete story of central black hole evolution.

\subsection{Easy-to-use relations}
\label{sec:easy-to-use}
Not all variables used in \cref{sec:low-scatter} are easy to observe in practice, therefore we here we restrict ourselves 35 different host galaxy properties that is easy to observe, and commonly used in most surveys. 
The second half of \cref{tab:relations} gives a set of ``easy-to-use" relations that can be applied to most of the galaxy surveys, with much lower RMSEs, as well as better performances on AIC \& BIC (\cref{tab:AICBIC}) than any of the old 2-D relations.

\section{Conclusions}
\label{sec:conclusion}

In this work, we have endeavored to produce meaningful black hole mass scaling relations that satisfy the needs of both theorist and observers.
The ``low-scatter'' relations can shed fresh light on how an SMBH co-evolves with its host galaxy. 
On the other hand, the ``easy-to-use" relations enable the best possible prediction of black hole masses for billions of galaxies that cannot have their central black hole masses ``directly" measured.
The extra precision in prediction power is invaluable for a better black hole mass function or any other demographic study of black holes, and will boost the discovery of IMBHs (intermediate-mass black holes).

Scaling relations are essentially empirically relations or effective models that work well in practice, and also hint some physics behind the correlated quantities.
SR is exactly the perfect tool to look for scaling relations due to its precision vs.\ simplistic nature.
Scaling relations are heavily used in nearly every sub-field within astrophysics. SR can power the discovery of meaningful higher-dimensional scaling relations and refresh our understanding of the Universe.

\begin{ack}
This material is based upon work supported by Tamkeen under the NYU Abu Dhabi Research Institute grant CASS.
\end{ack}

\bibliography{neurips}

\begin{thebibliography}{10}

\bibitem{blank_2019}
Marvin {Blank}, Andrea~V. {Macci{\`o}}, Aaron~A. {Dutton}, and Aura {Obreja}.
\newblock {NIHAO - XXII. Introducing black hole formation, accretion, and feedback into the NIHAO simulation suite}.
\newblock {\em \mnras}, 487(4):5476--5489, August 2019.

\bibitem{Akiyama_2019}
The Event Horizon~Telescope Collaboration, Kazunori Akiyama, Antxon Alberdi, Walter Alef, Keiichi Asada, Rebecca Azulay, Anne-Kathrin Baczko, David Ball, Mislav Baloković, John Barrett, Dan Bintley, Lindy Blackburn, Wilfred Boland, Katherine~L. Bouman, Geoffrey~C. Bower, Michael Bremer, Christiaan~D. Brinkerink, Roger Brissenden, Silke Britzen, Avery~E. Broderick, Dominique Broguiere, Thomas Bronzwaer, Do-Young Byun, John~E. Carlstrom, Andrew Chael, Chi kwan Chan, Shami Chatterjee, Koushik Chatterjee, Ming-Tang Chen, Yongjun Chen, Ilje Cho, Pierre Christian, John~E. Conway, James~M. Cordes, Geoffrey~B. Crew, Yuzhu Cui, Jordy Davelaar, Mariafelicia~De Laurentis, Roger Deane, Jessica Dempsey, Gregory Desvignes, Jason Dexter, Sheperd~S. Doeleman, Ralph~P. Eatough, Heino Falcke, Vincent~L. Fish, Ed~Fomalont, Raquel Fraga-Encinas, William~T. Freeman, Per Friberg, Christian~M. Fromm, José~L. Gómez, Peter Galison, Charles~F. Gammie, Roberto García, Olivier Gentaz, Boris Georgiev, Ciriaco Goddi, Roman Gold,
  Minfeng Gu, Mark Gurwell, Kazuhiro Hada, Michael~H. Hecht, Ronald Hesper, Luis~C. Ho, Paul Ho, Mareki Honma, Chih-Wei~L. Huang, Lei Huang, David~H. Hughes, Shiro Ikeda, Makoto Inoue, Sara Issaoun, David~J. James, Buell~T. Jannuzi, Michael Janssen, Britton Jeter, Wu~Jiang, Michael~D. Johnson, Svetlana Jorstad, Taehyun Jung, Mansour Karami, Ramesh Karuppusamy, Tomohisa Kawashima, Garrett~K. Keating, Mark Kettenis, Jae-Young Kim, Junhan Kim, Jongsoo Kim, Motoki Kino, Jun~Yi Koay, Patrick~M. Koch, Shoko Koyama, Michael Kramer, Carsten Kramer, Thomas~P. Krichbaum, Cheng-Yu Kuo, Tod~R. Lauer, Sang-Sung Lee, Yan-Rong Li, Zhiyuan Li, Michael Lindqvist, Kuo Liu, Elisabetta Liuzzo, Wen-Ping Lo, Andrei~P. Lobanov, Laurent Loinard, Colin Lonsdale, Ru-Sen Lu, Nicholas~R. MacDonald, Jirong Mao, Sera Markoff, Daniel~P. Marrone, Alan~P. Marscher, Iván Martí-Vidal, Satoki Matsushita, Lynn~D. Matthews, Lia Medeiros, Karl~M. Menten, Yosuke Mizuno, Izumi Mizuno, James~M. Moran, Kotaro Moriyama, Monika Moscibrodzka, Cornelia
  Müller, Hiroshi Nagai, Neil~M. Nagar, Masanori Nakamura, Ramesh Narayan, Gopal Narayanan, Iniyan Natarajan, Roberto Neri, Chunchong Ni, Aristeidis Noutsos, Hiroki Okino, Héctor Olivares, Gisela~N. Ortiz-León, Tomoaki Oyama, Feryal Özel, Daniel C.~M. Palumbo, Nimesh Patel, Ue-Li Pen, Dominic~W. Pesce, Vincent Piétu, Richard Plambeck, Aleksandar PopStefanija, Oliver Porth, Ben Prather, Jorge~A. Preciado-López, Dimitrios Psaltis, Hung-Yi Pu, Venkatessh Ramakrishnan, Ramprasad Rao, Mark~G. Rawlings, Alexander~W. Raymond, Luciano Rezzolla, Bart Ripperda, Freek Roelofs, Alan Rogers, Eduardo Ros, Mel Rose, Arash Roshanineshat, Helge Rottmann, Alan~L. Roy, Chet Ruszczyk, Benjamin~R. Ryan, Kazi L.~J. Rygl, Salvador Sánchez, David Sánchez-Arguelles, Mahito Sasada, Tuomas Savolainen, F.~Peter Schloerb, Karl-Friedrich Schuster, Lijing Shao, Zhiqiang Shen, Des Small, Bong~Won Sohn, Jason SooHoo, Fumie Tazaki, Paul Tiede, Remo P.~J. Tilanus, Michael Titus, Kenji Toma, Pablo Torne, Tyler Trent, Sascha Trippe,
  Shuichiro Tsuda, Ilse van Bemmel, Huib~Jan van Langevelde, Daniel~R. van Rossum, Jan Wagner, John Wardle, Jonathan Weintroub, Norbert Wex, Robert Wharton, Maciek Wielgus, George~N. Wong, Qingwen Wu, Ken Young, André Young, Ziri Younsi, Feng Yuan, Ye-Fei Yuan, J.~Anton Zensus, Guangyao Zhao, Shan-Shan Zhao, Ziyan Zhu, Juan-Carlos Algaba, Alexander Allardi, Rodrigo Amestica, Jadyn Anczarski, Uwe Bach, Frederick~K. Baganoff, Christopher Beaudoin, Bradford~A. Benson, Ryan Berthold, Jay~M. Blanchard, Ray Blundell, Sandra Bustamente, Roger Cappallo, Edgar Castillo-Domínguez, Chih-Cheng Chang, Shu-Hao Chang, Song-Chu Chang, Chung-Chen Chen, Ryan Chilson, Tim~C. Chuter, Rodrigo~Córdova Rosado, Iain~M. Coulson, Thomas~M. Crawford, Joseph Crowley, John David, Mark Derome, Matthew Dexter, Sven Dornbusch, Kevin~A. Dudevoir, Sergio~A. Dzib, Andreas Eckart, Chris Eckert, Neal~R. Erickson, Wendeline~B. Everett, Aaron Faber, Joseph~R. Farah, Vernon Fath, Thomas~W. Folkers, David~C. Forbes, Robert Freund, Arturo~I.
  Gómez-Ruiz, David~M. Gale, Feng Gao, Gertie Geertsema, David~A. Graham, Christopher~H. Greer, Ronald Grosslein, Frédéric Gueth, Daryl Haggard, Nils~W. Halverson, Chih-Chiang Han, Kuo-Chang Han, Jinchi Hao, Yutaka Hasegawa, Jason~W. Henning, Antonio Hernández-Gómez, Rubén Herrero-Illana, Stefan Heyminck, Akihiko Hirota, James Hoge, Yau-De Huang, C.~M.~Violette Impellizzeri, Homin Jiang, Atish Kamble, Ryan Keisler, Kimihiro Kimura, Yusuke Kono, Derek Kubo, John Kuroda, Richard Lacasse, Robert~A. Laing, Erik~M. Leitch, Chao-Te Li, Lupin C.-C. Lin, Ching-Tang Liu, Kuan-Yu Liu, Li-Ming Lu, Ralph~G. Marson, Pierre~L. Martin-Cocher, Kyle~D. Massingill, Callie Matulonis, Martin~P. McColl, Stephen~R. McWhirter, Hugo Messias, Zheng Meyer-Zhao, Daniel Michalik, Alfredo Montaña, William Montgomerie, Matias Mora-Klein, Dirk Muders, Andrew Nadolski, Santiago Navarro, Joseph Neilsen, Chi~H. Nguyen, Hiroaki Nishioka, Timothy Norton, Michael~A. Nowak, George Nystrom, Hideo Ogawa, Peter Oshiro, Tomoaki Oyama, Harriet
  Parsons, Scott~N. Paine, Juan Peñalver, Neil~M. Phillips, Michael Poirier, Nicolas Pradel, Rurik~A. Primiani, Philippe~A. Raffin, Alexandra~S. Rahlin, George Reiland, Christopher Risacher, Ignacio Ruiz, Alejandro~F. Sáez-Madaín, Remi Sassella, Pim Schellart, Paul Shaw, Kevin~M. Silva, Hotaka Shiokawa, David~R. Smith, William Snow, Kamal Souccar, Don Sousa, T.~K. Sridharan, Ranjani Srinivasan, William Stahm, Anthony~A. Stark, Kyle Story, Sjoerd~T. Timmer, Laura Vertatschitsch, Craig Walther, Ta-Shun Wei, Nathan Whitehorn, Alan~R. Whitney, David~P. Woody, Jan G.~A. Wouterloot, Melvin Wright, Paul Yamaguchi, Chen-Yu Yu, Milagros Zeballos, Shuo Zhang, and Lucy Ziurys.
\newblock First m87 event horizon telescope results. i. the shadow of the supermassive black hole.
\newblock {\em The Astrophysical Journal Letters}, 875(1):L1, apr 2019.

\bibitem{cranmer_2023}
Miles Cranmer.
\newblock Interpretable machine learning for science with pysr and symbolicregression.jl, 2023.

\bibitem{cranmer_2020}
Miles Cranmer, Alvaro Sanchez-Gonzalez, Peter Battaglia, Rui Xu, Kyle Cranmer, David Spergel, and Shirley Ho.
\newblock Discovering symbolic models from deep learning with inductive biases, 2020.

\bibitem{cranmer_2019}
Miles~D. Cranmer, Rui Xu, Peter Battaglia, and Shirley Ho.
\newblock Learning symbolic physics with graph networks, 2019.

\bibitem{Davis_2018}
Benjamin~L. {Davis}, Alister~W. {Graham}, and Ewan {Cameron}.
\newblock {Black Hole Mass Scaling Relations for Spiral Galaxies. II. M $_{BH}$-M $_{*,tot}$ and M $_{BH}$-M $_{*,disk}$}.
\newblock {\em \apj}, 869(2):113, December 2018.

\bibitem{Davis_2019}
Benjamin~L. {Davis}, Alister~W. {Graham}, and Ewan {Cameron}.
\newblock {Black Hole Mass Scaling Relations for Spiral Galaxies. I. M $_{BH}$-M $_{*,sph}$}.
\newblock {\em \apj}, 873(1):85, March 2019.

\bibitem{Davis_2019b}
Benjamin~L. {Davis}, Alister~W. {Graham}, and Fran{\c{c}}oise {Combes}.
\newblock {A Consistent Set of Empirical Scaling Relations for Spiral Galaxies: The (v $_{max}$, M $_{oM}$)-({\ensuremath{\sigma}} $_{0}$, M $_{BH}$, {\ensuremath{\phi}}) Relations}.
\newblock {\em \apj}, 877(1):64, May 2019.

\bibitem{Davis_2017}
Benjamin~L. {Davis}, Alister~W. {Graham}, and Marc~S. {Seigar}.
\newblock {Updating the (supermassive black hole mass)-(spiral arm pitch angle) relation: a strong correlation for galaxies with pseudobulges}.
\newblock {\em \mnras}, 471(2):2187--2203, October 2017.

\bibitem{Davis_Jin_2023}
Benjamin~L. {Davis} and Zehao {Jin}.
\newblock {Discovery of a Planar Black Hole Mass Scaling Relation for Spiral Galaxies}.
\newblock {\em \apjl}, 956(1):L22, October 2023.

\bibitem{Graham_2022}
Alister~W Graham and Nandini Sahu.
\newblock {Appreciating mergers for understanding the non-linear Mbh–M*,spheroid and Mbh–M*, galaxy relations, updated herein, and the implications for the (reduced) role of AGN feedback}.
\newblock {\em Monthly Notices of the Royal Astronomical Society}, 518(2):2177--2200, 11 2022.

\bibitem{Graham_2013}
Alister~W. {Graham} and Nicholas {Scott}.
\newblock {The M $_{BH}$-L $_{spheroid}$ Relation at High and Low Masses, the Quadratic Growth of Black Holes, and Intermediate-mass Black Hole Candidates}.
\newblock {\em \apj}, 764(2):151, February 2013.

\bibitem{Kormendy_2013}
John {Kormendy} and Luis~C. {Ho}.
\newblock {Coevolution (Or Not) of Supermassive Black Holes and Host Galaxies}.
\newblock {\em \araa}, 51(1):511--653, August 2013.

\bibitem{Krone_Martins_2014}
A.~Krone-Martins, E.~E.~O. Ishida, and R.~S. de~Souza.
\newblock The first analytical expression to estimate photometric redshifts suggested by a machine.
\newblock {\em Monthly Notices of the Royal Astronomical Society: Letters}, 443(1):L34--L38, jun 2014.

\bibitem{lemos_2022}
Pablo Lemos, Niall Jeffrey, Miles Cranmer, Shirley Ho, and Peter Battaglia.
\newblock Rediscovering orbital mechanics with machine learning, 2022.

\bibitem{Nelson_2017}
Dylan Nelson, Annalisa Pillepich, Volker Springel, Rainer Weinberger, Lars Hernquist, Rüdiger Pakmor, Shy Genel, Paul Torrey, Mark Vogelsberger, Guinevere Kauffmann, Federico Marinacci, and Jill Naiman.
\newblock {First results from the IllustrisTNG simulations: the galaxy colour bimodality}.
\newblock {\em Monthly Notices of the Royal Astronomical Society}, 475(1):624--647, 11 2017.

\bibitem{Onorbe_2015}
Jose {O{\~n}orbe}, Michael {Boylan-Kolchin}, James~S. {Bullock}, Philip~F. {Hopkins}, Du{\v{s}}an {Kere{\v{s}}}, Claude-Andr{\'e} {Faucher-Gigu{\`e}re}, Eliot {Quataert}, and Norman {Murray}.
\newblock {Forged in FIRE: cusps, cores and baryons in low-mass dwarf galaxies}.
\newblock {\em \mnras}, 454(2):2092--2106, December 2015.

\bibitem{Robotham_2015}
A.~S.~G. {Robotham} and D.~{Obreschkow}.
\newblock {Hyper-Fit: Fitting Linear Models to Multidimensional Data with Multivariate Gaussian Uncertainties}.
\newblock {\em \pasa}, 32:e033, September 2015.

\bibitem{Sahu_2019b}
Nandini {Sahu}, Alister~W. {Graham}, and Benjamin~L. {Davis}.
\newblock {Black Hole Mass Scaling Relations for Early-type Galaxies. I. M $_{BH}$-M $_{*,}$ $_{sph}$ and M $_{BH}$-M $_{*,gal}$}.
\newblock {\em \apj}, 876(2):155, May 2019.

\bibitem{Sahu_2019}
Nandini {Sahu}, Alister~W. {Graham}, and Benjamin~L. {Davis}.
\newblock {Revealing Hidden Substructures in the M $_{BH}$-{\ensuremath{\sigma}} Diagram, and Refining the Bend in the L-{\ensuremath{\sigma}} Relation}.
\newblock {\em \apj}, 887(1):10, December 2019.

\bibitem{Sahu_2020}
Nandini Sahu, Alister~W. Graham, and Benjamin~L. Davis.
\newblock Defining the (black hole)–spheroid connection with the discovery of morphology-dependent substructure in the mbh–nsph and mbh–re,sph diagrams: New tests for advanced theories and realistic simulations.
\newblock {\em The Astrophysical Journal}, 903(2):97, nov 2020.

\bibitem{Sahu_2022}
Nandini {Sahu}, Alister~W. {Graham}, and Benjamin~L. {Davis}.
\newblock {The (Black Hole Mass)-(Spheroid Stellar Density) Relations: M $_{BH}$-{\ensuremath{\mu}} (and M $_{BH}$-{\ensuremath{\Sigma}}) and M $_{BH}$-{\ensuremath{\rho}}}.
\newblock {\em \apj}, 927(1):67, March 2022.

\bibitem{Savorgnan_2016}
G.~A.~D. {Savorgnan} and A.~W. {Graham}.
\newblock {Supermassive Black Holes and Their Host Spheroids. I. Disassembling Galaxies}.
\newblock {\em \apjs}, 222(1):10, January 2016.

\end{thebibliography}
\bibliographystyle{plain}

\newpage
\appendix
\label{sec:appendix}

\begin{table*}[t]
\tiny
\caption{AIC and BIC score comparison between SR-found N-D relations and old 2-D relations. A lower AIC or BIC indicates a model that has a better balance between goodness of fit and complexity.}
\label{tab:AICBIC}
\vskip 0.15in
\begin{center}
\begin{tabular}{lll}
\toprule
 & \textsc{AIC} & \textsc{BIC} \\
\toprule
\textbf{Low-scatter relation} \\
\midrule
$2.85 \log \left(\frac{\sigma_0}{189}\right) + 1.16 \log \left(\frac{B/T}{0.437}\right) - 0.33 \log \left(\frac{\rho_\text{soi}}{601}\right) + 8.20 $ 
& 713 & 722 \\

\toprule
\textbf{Easy-to-use relations} \\
\midrule

$\log \sigma_0 + \log \textup{M}^{*}_{\textup{sph}} - 0.56 \ \textup{Pseudobulge} - 4.57$
& 791 & 799 \\

\hline

$3.59 \log \sigma_0 + 0.50 \log R_{e,sph,maj} - 0.50 \textup{Pseudobulge}$
& 770 & 778 \\

\toprule
\textbf{2-D relations} \\
\midrule

$6.10 \log (\sigma_0/200)+8.27$
& 980 & 982 \\

\hline

$1.31 \log \text{M}^*_\textup{sph} - 5.83$
& 859 & 862 \\

\hline

$2.40 \log(B/T) + 9.11$
& 921 & 924 \\

\hline

$-0.93 \log \rho_\textup{soi} + 10.64$
& 951 & 954 \\

\bottomrule
\end{tabular}
\end{center}
\vskip -0.1in
\end{table*}

\begin{figure*}[hb]
\vskip 0.2in
\begin{center}
\includegraphics[width=0.31\linewidth]{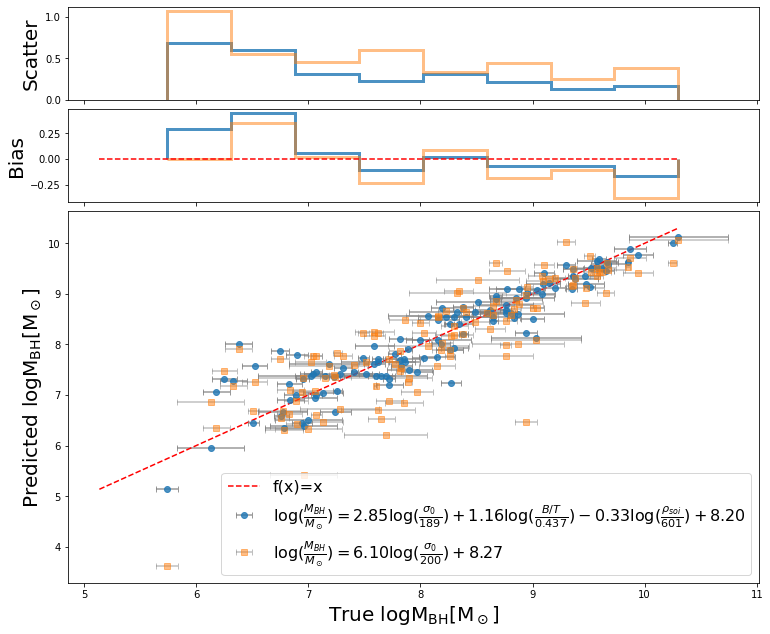}
\includegraphics[width=0.31\linewidth]{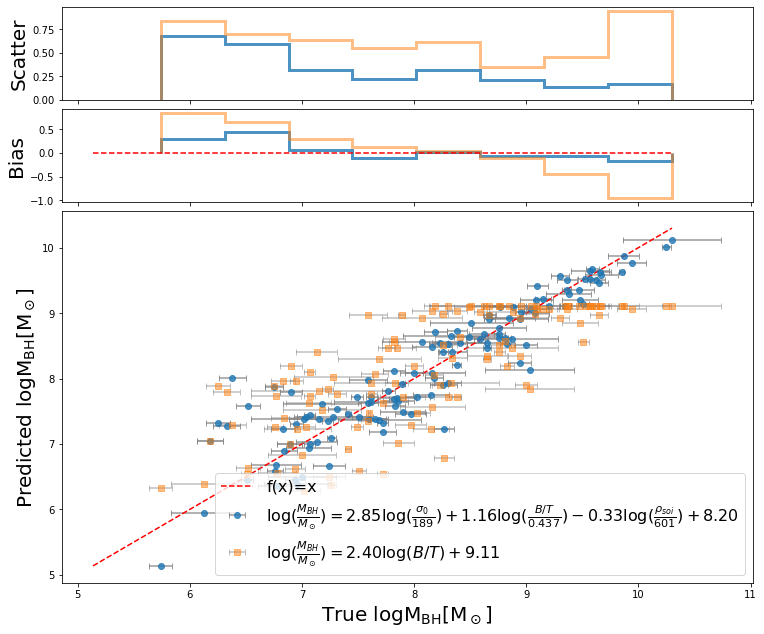}
\includegraphics[width=0.31\linewidth]{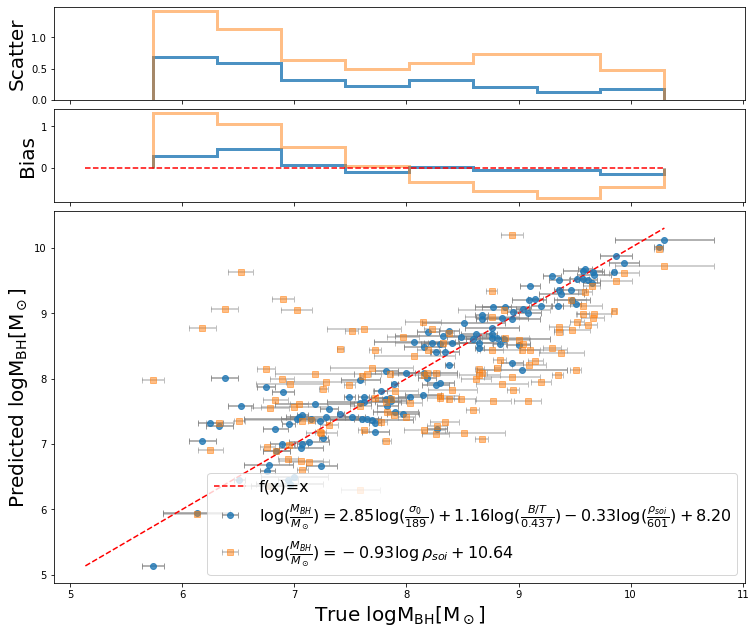}
\includegraphics[width=0.31\linewidth]{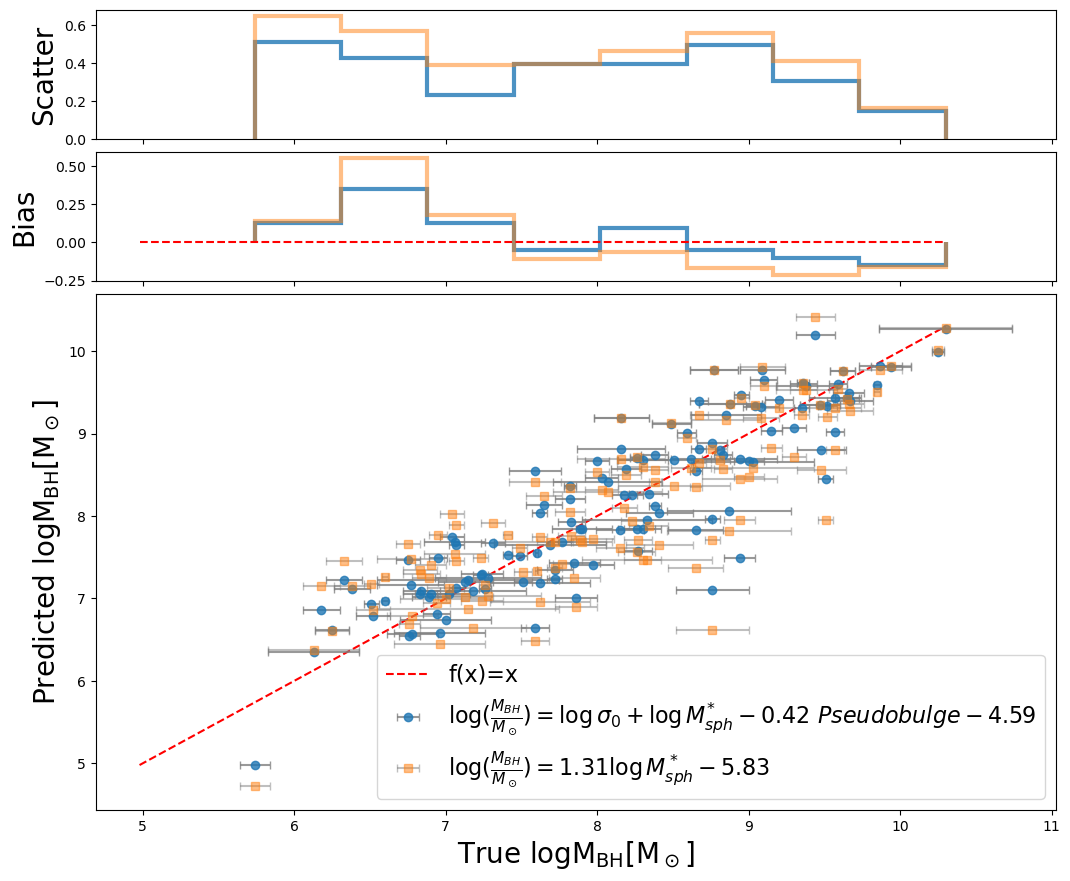}
\includegraphics[width=0.31\linewidth]{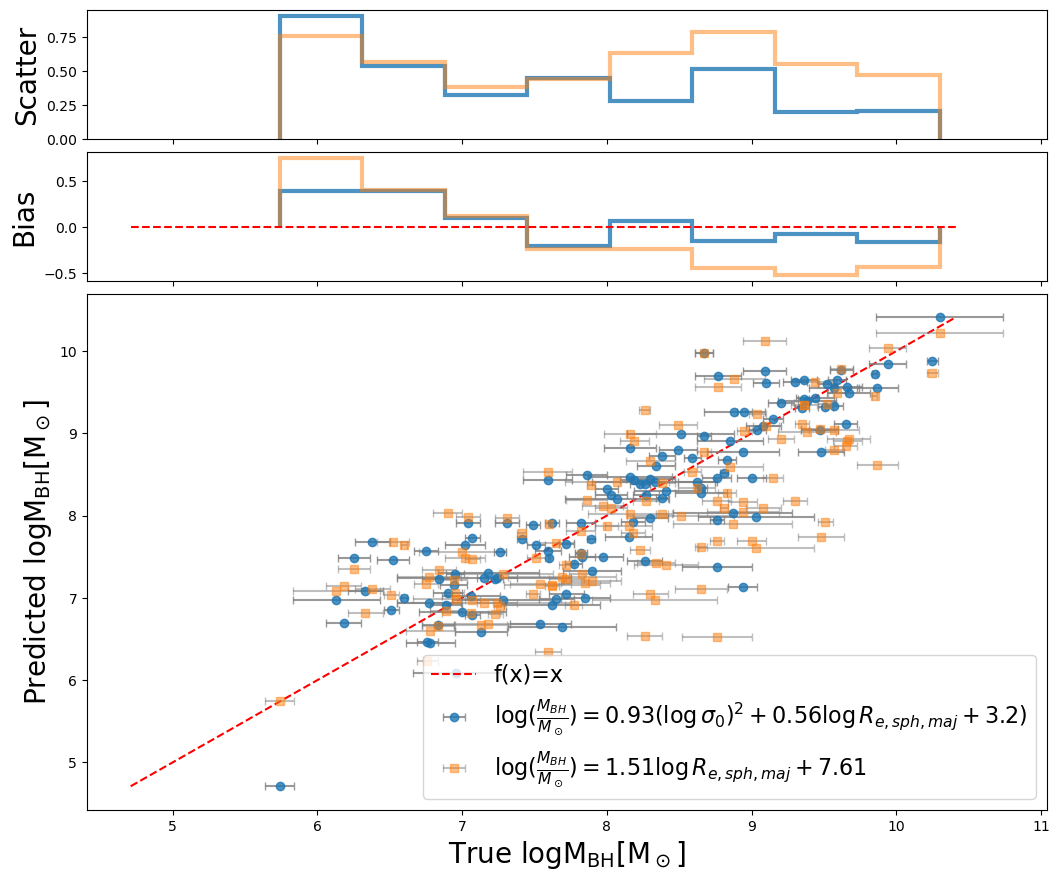}
\includegraphics[width=0.31\linewidth]{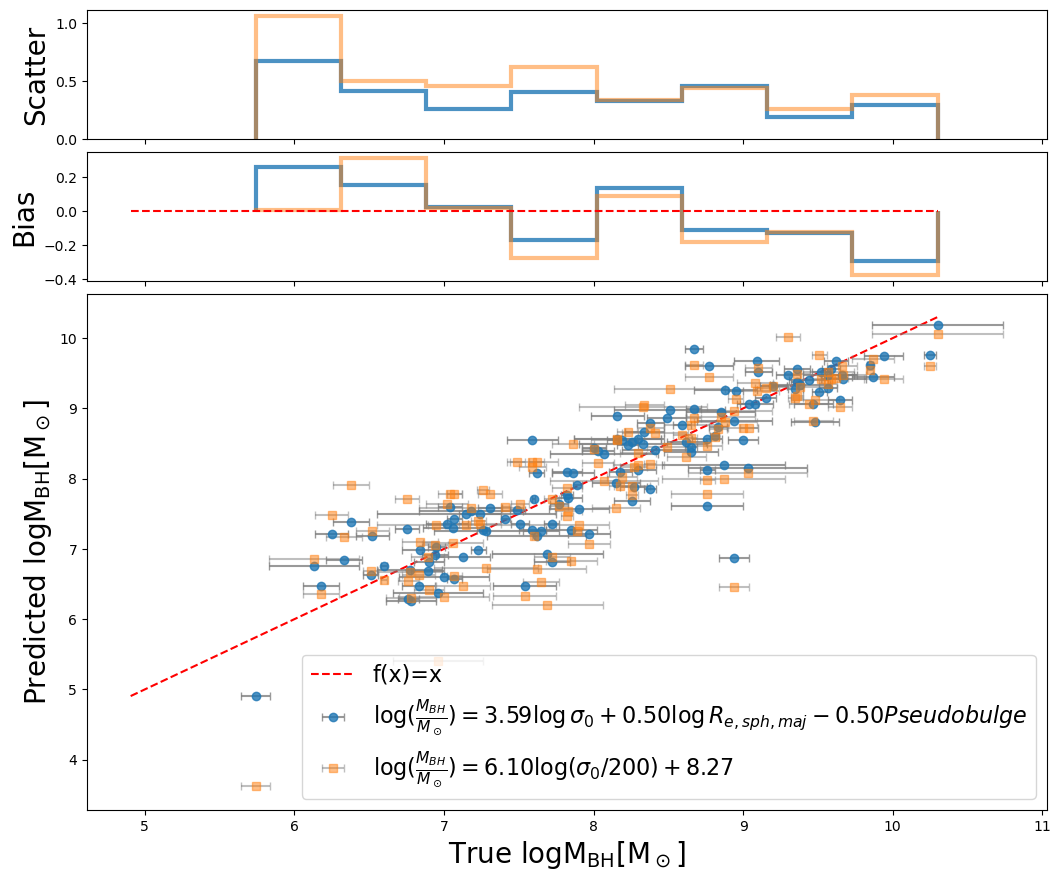}
\caption{N-D black hole mass scaling relations (blue) compared with their 2-D counterpart relations (orange).
The red dotted line represents where predicted black hole mass perfectly equals the true black hole mass, thus a point closer to the red rotted line means better prediction.
The ``scatter" panels show the average residual ($\sqrt{(\text{Truth}-\text{Prediction})^2}$) across different black hole mass range bins; while the ``bias" panels show the average difference (\text{Truth}-\text{Prediction}) across different mass range bins to check if a relation tends to over-predict or under-predict black hole mass.
The 1$^\text{st}$ row is devoted to one of the best ``low-scatter" relations \cref{equ:best}, compared with its three 2-D counterparts.
The 2$^\text{nd}$ row shows ``easy-to-use" relations. }
\label{fig:relations}
\end{center}
\vskip -0.2in
\end{figure*}


\end{document}